\documentclass[12pt,preprint]{aastex}

\newcommand{\udeg}{\hbox{$^\circ$}}

\shorttitle{The Evolution of Tully-Fisher at $z\approx0.9$}
\shortauthors{Barden et al.}

\begin{document}

\title{H$\alpha$ Rotation Curves of $z\sim 1$ Galaxies: \\ Unraveling
the Evolution of the Tully-Fisher Relation\altaffilmark{1}}

\author{M. Barden\altaffilmark{2}, M. D. Lehnert\altaffilmark{2},
L. Tacconi\altaffilmark{2}, R. Genzel\altaffilmark{2},
S. White\altaffilmark{3}, \& A. Franceschini\altaffilmark{4}}

\altaffiltext{1}{Based on observations collected at the European Southern
Observatory, Cerro Paranal, Chile}

\altaffiltext{2}{Max-Planck-Institut f\"ur extraterrestrische Physik,
Giessenbachstra\ss e, 85748 Garching bei M\"unchen, Germany}

\altaffiltext{3}{Max-Planck-Institut f\"ur Astrophysik, 
Karl-Schwarzschild Str. 1, 85748 Garching bei M\"unchen, Germany}

\altaffiltext{4}{Dipartimento di Astronomia, Universita di Padova,
Vicolo Osservatorio 2, I-35122, Italy}

\begin{abstract}
We describe the first results of a programme to obtain rotation curves
of $z\approx 1$ disc galaxies in the near-infrared using the
H$\alpha$ emission line in order to study the Tully-Fisher relationship.  
To put any observed evolution into perspective and to investigate any possible 
selection biases, we constructed a control sample of low redshift 
galaxies that had rotation velocities and images available for measuring 
their dynamical, photometric, and morphological properties.
Compared to local objects with isophotal sizes similar to the high 
redshift targets, we find that our sample of galaxies with spatially 
resolved rotation curves, the most 
distant sample so far ($\left\langle z \right\rangle \approx 0.9$), clearly
reveals a brightening of $\approx 1.1$ mag in the rest-frame B-band.
The observed offset can be explained by a combination of increasing 
surface brightness, decreasing rotation speeds, and slightly smaller 
disc scale lengths of the high redshift galaxies.
\end{abstract}


\section{Introduction}
Like the fundamental plane for early type galaxies 
\citep{1987ApJ...313...42D} the Tully-Fisher 
relation for disc galaxies \citep{1977A&A....54..661T} embodies fundamental 
implications for the
relationship between the mass of the galaxy, its star-formation history,
specific angular momentum and dark matter content and distribution.

Broadly speaking, there are two competing models to
explain the Tully-Fisher relation. The first of these is that it 
is a consequence of self-regulated star formation
in discs with different masses \citep[e.g.,][]{1997ApJ...481..703S},
i.e., the competition of disc instability (which promotes star
formation) with supernovae induced porosity (which inhibits star-formation).  
The model is not the complete answer, however,
since it does not explain the mass-to-light ratios or
the scale-lengths of the discs. In the second 
model the Tully-Fisher relation is a direct consequence of
the cosmological equivalence between mass and circular velocity
\citep[e.g.,][]{1998MNRAS.295..319M, 1999ApJ...513..555S}.  This
formalism is part of what has become the standard model for the growth
of structure - the hierarchical merging model in which the gravitational
effects of dark matter drive the evolution of galaxies and large-scale
structure \citep[e.g.,][]{1993MNRAS.264..201K}. Models of this type
have the advantage of providing testable predictions about the sizes,
surface densities, and rotation curves of galaxies as a function of
redshift. However, as emphasized by \citet{1999ApJ...513..555S}, although
the Tully-Fisher relation can naturally be explained by hierarchical
merging models, the normalization and evolution of the Tully-Fisher
relation depend strongly on the prescription used for the star formation 
and on the cosmological parameters.

It is now well established that massive disc galaxies exist out to
redshifts $z\sim 1$ \citep{1996ApJ...465L..15V,1997ApJ...479L.121V,
1996ApJ...465L.103S, 1998ApJ...500...75L}. For a given size scale, the
number density of these disc galaxies is approximately the same at $z\sim
0.8$ as is observed locally. Overall, the results at moderate redshift
($z\sim 0.5$) are rather mixed. Depending on the sample selection, the
technique used to estimate the rotation speed, the median redshift of the
sample, and the wavelength at which comparisons are made, there are claims
in the literature that the Tully-Fisher relation either brightens or dims
with redshift \citep[see e.g.,][]{1996ApJ...465L..15V,1997ApJ...479L.121V,
1998ApJ...505...96S, 1997MNRAS.285..779R, 2002ApJ...564L..69Z}. To
help resolve this controversy and to push to higher redshift where
evidence for evolution of disc galaxies will likely be stronger
and more obvious, we set out to obtain spatially resolved rotation
curves at the highest redshifts where large samples are available.
Obtaining spatially resolved rotation curves becomes quite difficult at
such high redshifts since [OII]$\lambda$3727 is redshifted into a region
with many strong night sky lines  and the sensitivity of CCDs declines
rapidly. Instead we have chosen to observe the H$\alpha$ line, reshifted
to the infrared J-band. H$\alpha$ is typically 2.5 times stronger than
[OII] \citep{1992ApJ...388..310K} and being at longer wavelength, is
less affected by dust extinction.

\section{Sample Selection}

To gauge the evolution of the Tully-Fisher relation we compare a
local with a high redshift sample of highly inclined, ``normal''
spiral galaxies. The high redshift sample consists of objects with
measured spectroscopic redshifts. To be able to measure H$\alpha$
in the near-infrared we required $z>0.6$.  Targets were selected
from the CFRS / LDSS redshift surveys, the clusters MS1054, AC103
and the Hawaii deep field SSA22 \citep[][respectively, and references
therein]{1995ApJ...455...50L, 1995MNRAS.273..157G, vandokkumthesis,
1998ApJ...497..188C, 1996AJ....112..839C}.  Furthermore, we included
two targets from the VLT science archive.  For the majority of these
sources we have HST I-band\footnote{Based on observations made with
the NASA/ESA Hubble Space Telescope, obtained from the data archive at
the Space Telescope Institute. STScI is operated by the association of
Universities for Research in Astronomy, Inc. under the NASA contract
NAS 5-26555.} or at least VLT J-band images. For all of them we have
obtained H$\alpha$ ISAAC long-slit spectra to determine spatially resolved
rotation curves.  For this purpose we have selected these sources to be
spatially extended on the images (diameters mainly $>$2'') and to have
high inclination angles ($i>30\udeg$, $\left\langle i \right\rangle
\approx60\udeg$). Furthermore, we chose only objects with exponential
light profiles to ensure that we were observing disc galaxies.  The mean
redshift of our high-z sample is $z\approx0.92$ (see Tab.~\ref{tbl-1}).

To construct a complementary sample of local galaxies we searched the
literature for objects with accessible CCD B-band images (obtained
from the NASA Extragalactic Database, NED, or ESO science archive),
published distances based on primary distance indicators or mean group
velocities \citep[][ LEDA database\footnote{The Lyon Meudon extragalactic
database (\url{http://leda.univ-lyon1.fr)}}]{2001ApJ...553...47F,
1992ApJ...387...47P, 1988AJ.....95.1025P, 2000ApJ...533..744T,
2000ApJS..128..431F, 2000ApJ...543..178G, 2000A&AS..142..425D,
1997A&A...324..457K, 2000A&A...362..544K, 1990A&AS...86....7D,
1989ApJS...70....1A, 1985A&AS...61..141D} and an estimate for the
rotation speed based on either an HI or a resolved H$\alpha$ rotation
curve \citep[][ and references therein]{1998A&AS..127..117P}. 
Total magnitudes for all local sample galaxies where obtained
from RC3 \citep{1991RC3...C......0D} or LEDA. As a consistency check,
the object magnitudes, sizes, rotation speeds and the resulting
Tully-Fisher relation were compared to the RC3 catalogue and the data
from \citet{2000ApJ...533..744T}.  From this comparison, we conclude
that our sample of low redshift galaxies is in good agreement with a
random subsample of a magnitude limited complete sample from the RC3.
      
\section{Observations}

We obtained J- and H-band spectroscopy during four observing runs with the
facility near-infrared spectrometer ISAAC at the VLT 
\citep{1997SPIE.2871.1146M,1998Msngr..94....7M}. We used the medium resolution
spectroscopic mode with a slit-width resolution product of 
$R\approx 3000-5000$ and a spatial pixel scale of 0.1484''.

In ESO-period 65 and 66 (April - September 2000 and October 2000 -
March 2001 respectively) we obtained the first set of data in service mode.
With total integration times of 3 hours and a
0.6''-slit ($R\approx 5200$), which we rotated according to the position
angle of the objects, we could successfully observe four targets.
The target selection and successive estimation of position angles
for this set we performed on 1'' resolution images acquired in the J-band
with ISAAC (20min integration time) before the spectroscopic observations.
In addition to these J-band images, we obtained I-band images taken with
the Canada France Hawaii Telescope (CFHT) from that telescope's archive.

Furthermore, we conducted three other runs in visitor mode in February
and September 2001 and September 2002. This time the set-up included
a 1''-slit ($R\approx 3100$) and integration times varied, depending
on the H$\alpha$ flux of the targets, between 1 and 3 hours. Moreover,
we selected targets on the basis of HST/WFPC2 F814W images, which we
also used to orient the slit along the galaxy major axis. Since most
of the targets were too faint to be acquired directly we used a blind
off-set from a``bright'' star, calculated from the HST images. Using
this prescription we observed another 17 objects successfully. The seeing
during these four runs ranged from 0.4'' to 0.8''.

\section{Data Reduction and Analysis}

To reduce the ISAAC spectra we used standard data reduction packages
(ESO eclipse \& IRAF). The data were flat fielded and dark-subtracted.
After dark-subtraction and cosmic ray rejection we checked all frames
individually for shifts in the wavelength direction due to flexure and
applied a reverse shift if necessary. Before combining the frames we
calculated a wavelength solution and distortion correction. Finally,
we subtracted a background and smoothed the spectra with a Gaussian of
FWHM approximately equal to the number of pixels per resolution element.

Firstly, we extracted one dimensional H$\alpha$-profiles 
across the spectral and spatial regions of the array that
contained H$\alpha$-flux. Typically, even with the increased 
signal-to-noise gained in the continuum by summing up the flux in this 
way, we could not detect the continuum significantly. We then fitted the 
one dimensional H$\alpha$-profiles (typically assuming no galaxy continuum) 
by one or two (in the case of a double horned profile) Gaussians. At the 
level where $20\%$ of the total integrated flux was reached we measured 
the width of the profile. Applying the corrections given by 
\citet{1985ApJS...58...67T} we obtained a first estimate for the terminal 
rotation velocity of our sample galaxies. 

\begin{figure}
\plotone{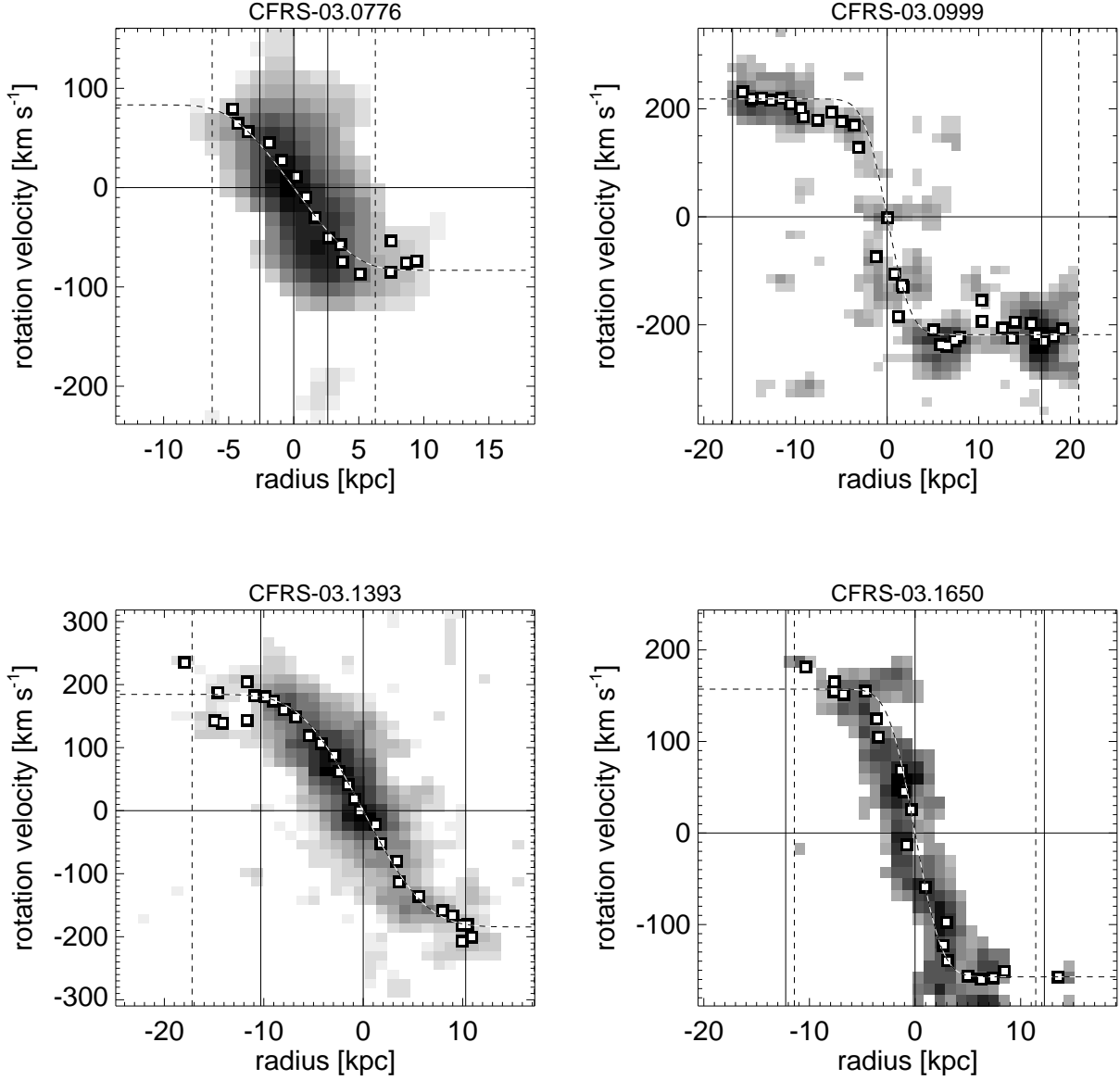}
\caption{Typical 2-D spectra for some galaxies of our high redshift 
sample. The
grey-scale shows the 2-D H$\alpha$ spectra, the squares represent the
adopted position-velocity measurements with the best-fit model rotation curve
over-layed as a dashed line, and the vertical dashed and/or solid
lines show the isophotal radius $R_{25}$ at a surface brightness of 25 
mag arcsec$^{-2}$ and 2.2 times the disc scale length 
$R_{2.2}=2.2 \times R_d$, respectively.
\label{rotation-profile}}
\end{figure}

In all cases but one, however, we could derive a terminal rotation
velocity from a resolved rotation curve model. To extract the data points
for the model rotation curve we developed a special adaptive curve tracing
algorithm.  We then fitted the resulting position-velocity diagram with
a simple model of a flat rotation curve (which is appropriate for not
too-late disc galaxies). This step-function we convolved with a Gaussian
with a full-width at half maximum of the seeing. Moreover, the convolution
included weighting with an exponential function with a scale-length as
measured from our disc surface brightness fits on the HST or VLT images
assuming that line and continuum have the same light distribution. Tests
show that our modeling delivers robust rotation velocities, which
are not very sensitive to the exact weighting scheme, the impact of
nuclear bulges, the signal-to-noise ratio and in particular, how well
the observationally ``flat part'' of the rotation curve is sampled by
the data. We have measured the seeing from the J-band acquisition images
taken before the observations, the acquisition images of a subsequent
target, and the values from the visible seeing monitor. Given our long
integration times and the limited sampling of the seeing at the observing
wavelength, the seeing is somewhat uncertain. We have therefore let it
be a free parameter in the fitting process. The values obtained from an
unconstrained fit correlate well with the seeing values estimated from the
acquisition images and seeing monitor, though. Some of the final fits to
the data are shown in Fig.~\ref{rotation-profile}. Although only $50\%-60\%$
of the data show a clear turn-over or extend out to 2.2 disc scale lengths
\citep[which is were the rotation curve of a self-gravitating thin disc 
reaches its maximum, see ][]
{1987gady.book.....B} the remaining data do not exclusively consist
of slow rotators only. In fact, the distributions of rotation speeds
of high and low quality data can not be distinguished from each other
(see Barden et al. 2003).
Furthermore, the rotation speeds derived from the two dimensional data are
not significantly smaller than the ones from the one dimensional spectra,
which have much higher signal-to-noise (see Barden et al. 2003). 
Moreover, according to the parametric fits to over 2000 local disc galaxies of 
\citet{2002ApJ...571L.107G} at radii varying from $1...1.5\times R_d$ one 
should already measure at least $\sim 80\%$ of the asymptotic rotation 
velocity. Since the majority of our high galaxies extend further out we 
estimate rotation velocities within $<20\%$.
Therefore, we are confident that we do not underestimate the rotation 
velocities significantly. Finally, the rotation
velocities have to be corrected for the inclination of the galaxies. We
have derived inclinations from surface brightness profile fits to our
HST or ISAAC images.

To estimate rest-frame B-band magnitudes we compiled all available
photometric information for our sources (two to five independent
measurements). For each magnitude we calculated the corresponding
rest-frame wavelength and corrected the magnitudes for galactic foreground
extinction \citep{1998ApJ...500..525S}. Then we fitted the extinction
corrected magnitudes and corresponding rest-wavelengths with spectra
from the atlas of \citet{1996ApJ...467...38K}. We note that our final
rest-frame B-band magnitudes are relatively insensitive to the models
used since all of the galaxies had a photometric band available that was
very near to the B-band rest wavelength. Finally, we corrected the 
resulting rest-frame B-band magnitudes for inclination dependent internal 
extinction according to the method of \citet{1985ApJS...58...67T} and 
determined a face-on B-magnitude (Table~\ref{tbl-1}).

The parameters for the local comparison sample were determined and
corrected in exactly the same manner as for the high redshift galaxies.
Rotation speeds were extracted from HI or optical rotation curves.
Only one object was included where a spatially-resolved rotation curve
was not available. Comparison of the resolved data with the HI line
width from RC3 usually agree within $\pm 20$ km s$^{-1}$ (see Barden et
al. 2003). We calculated inclination corrections based on inclinations
from our own profile fits in agreement with the values from RC3 or
LEDA. Total magnitudes were obtained directly from RC3 or LEDA and were
corrected for galactic foreground extinction and internal extinction.
Since differences in the isophotal extent between the images of the local
and high redshift galaxies can be quite dramatic (up to two magnitudes),
the light profiles of the local galaxy sample were fit only out to
a characteristic limiting radius, which was approximately the average
isophotal depth of the images of the high redshift galaxies (taking into
account cosmic surface brightness dimming; see Barden et al. 2003).

\section{The Tully-Fisher Relation at $z\approx 0.9$}

\begin{figure}
\plotone{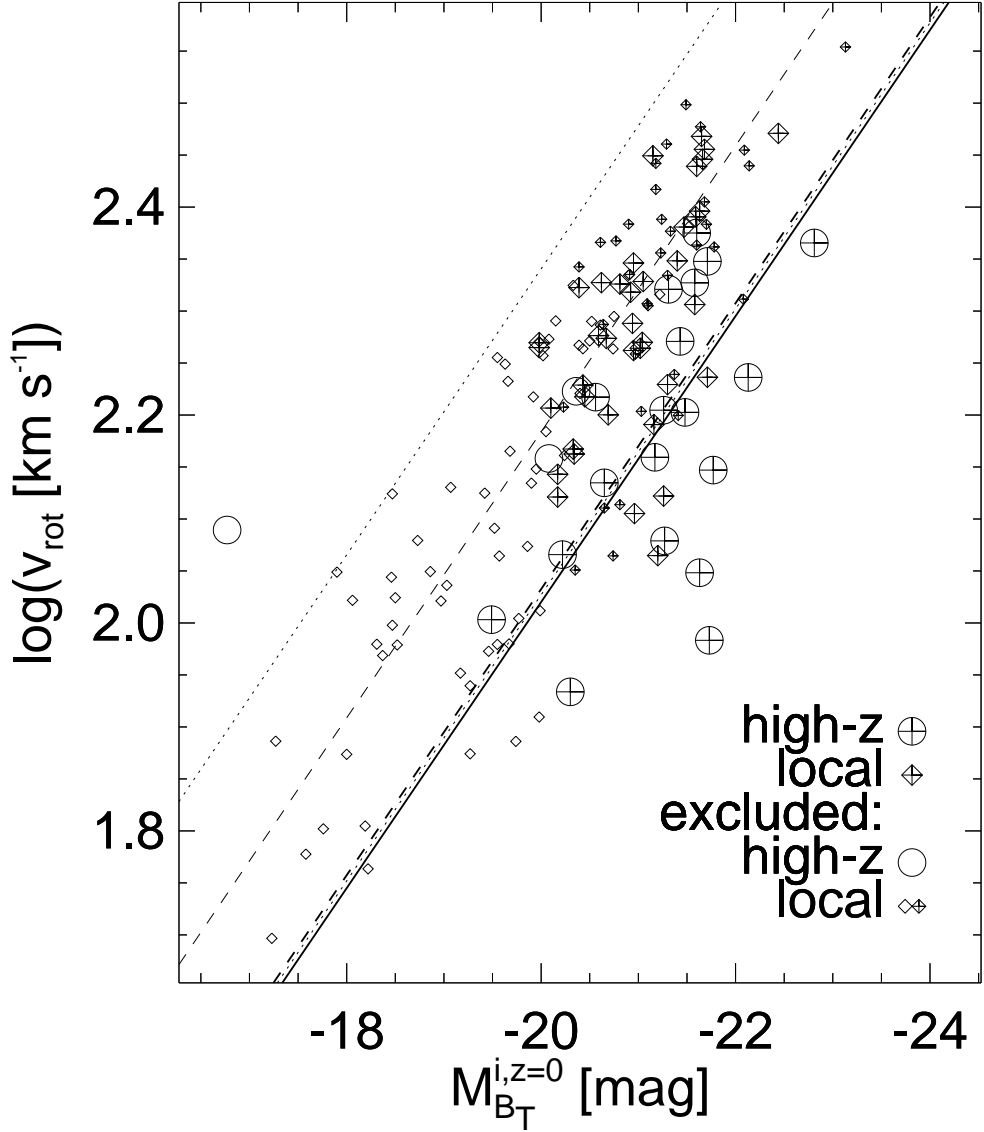}
\caption{Tully-Fisher relation for the high and low redshift samples. 
Circles correspond to high redshift galaxies, diamonds to local galaxies. 
Pluses mark isophotal sizes larger than the required minimum 
$R_{25}>12.5$ kpc. Only local galaxies shown as large symbols are included 
in the control sample in order to match the RC3 statistics. 
The thin lines mark the local relation by 
\citet{2000ApJ...533..744T} (thin dashed line) and its $\pm3\sigma$ 
error bars (thin dotted lines). A fit to the high redshift data results 
in a Tully-Fisher relation -1.1mag offset from the local Tully-Fisher 
(thick solid line). Our reconstruction
is plotted as the thick dashed line. The agreement between fit and our 
high redshift calibration is astonishing.
\label{tully-fisher}}
\end{figure}

Comparing the isophotal-size-culled local (our comparison sample) and
high redshift galaxy samples (see Barden et al. 2003) we find an offset
for the high-z data from the Tully-Fisher relation of $\Delta M=-1.1\pm
0.2$ magnitudes in a $\Lambda$-dominated universe ($H_{0}=70$, $\Omega
_{M}=0.3$, $\Omega _{\Lambda }=0.7$; Fig.~\ref{tully-fisher}). This
``brightening'' of the distant galaxies is significant and holds for
any reasonable cosmology ($\Delta M=-0.5\pm 0.2$ for $\Omega _{M}=1$
and $\Omega _{\Lambda }=0$; $\Delta M=-1.0\pm 0.2$ for $\Omega _{M}=0.05$
and $\Omega _{\Lambda }=0$). The limited range in isophotal 
size makes measuring a slope of the Tully-Fisher relation impossible.

\section{Discussion and Conclusions}

The observed offset in the Tully-Fisher relation at $\left\langle z
\right\rangle\approx 0.9$ for large galaxies is partially a consequence
of the fact that we observe an increase of central surface brightness of
1.7 mag arcsec$^{-2}$ and a slight decrease of $\approx 30\%$ in the disc
scale lengths of the high redshift sample compared to the local control
sample (see Barden et al. 2003). Using the relationship between disc
magnitude, central surface brightness, and disc scale length, namely,
$M_d\propto-5\log R_d-2.5\log \left(1+z\right)+\mu_0-const$, we find
that the differences in average central surface brightness and average
disc scale lengths imply an offset in the zero-point of the T-F relation
between the two samples of about $-0.3\pm0.4$ magnitudes.  In addition, by
comparing the rotation speeds of the high redshift galaxies and the local
comparison sample, we find an average offset of $\Delta v_{rot}=50\pm 15$
km s$^{-1}$ of the high redshift sample (see Barden et al. 2003).  Thus,
between $z=0.9$ and 0 the mean mass of galaxies has increased by a factor
of $2.6\pm 0.4$ (assuming mass $\propto\, v_{rot}^2$, and estimating
the total mass out to 2.2 disc scale lengths measured from fitting
the light profile).  Therefore, if we use a parameterization of the
Tully-Fisher relation as given by \citet{2000ApJ...533..744T}, adjusting
the zero-points by -0.3 mag to account for the average differences in
surface brightness and disc scale length and decreasing the average
rotation speed by $25\%$ yields an offset from the local Tully-Fisher
relation consistent with that observed ($\Delta M=-1.1\pm 0.5$ mag).

Having the highest redshift sample puts us in a unique position to discuss
the evolution of the Tully-Fisher relation from a new perspective. All of
the studies to date have found a significantly smaller amount of evolution
compared to the results presented here. The most obvious way to reconcile
this difference is that the offset in the Tully-Fisher relation grows
with increasing redshift. The studies with the highest median redshift
currently are \cite{2002ApJ...564L..69Z} and \cite{1996ApJ...465L..15V}
with a mean redshift of $\left\langle z \right\rangle \sim 0.4-0.5$,
compared with our mean redshift of $\left\langle z \right\rangle
\approx 0.9$. The time difference is therefore about 2-3 Gyrs for
any reasonable cosmology, clearly sufficient for galaxies to undergo
significant evolution.

Unfortunately, comparing these results with model predictions is not very
straight-forward since the exact offset predicted depends on the assumed 
physics and phenomenology, such as the details of star-formation 
\citep[e.g.,][]{1999ApJ...513..555S,
1998MNRAS.295..319M}.  Recently, within the frame work of galaxy evolution
\citep[e.g.,][]{1978MNRAS.183..341W}, a number of authors have made 
predictions about the general evolution of the galaxy
population.  For example, \citet{2001MNRAS.327L..10N} predict that from
$z=0$ to 1, using the cosmological parameters we have assumed here, that the
overall luminosity function of galaxies in the B-band will brighten by
about 1.15 mag and that the characteristic mass of galaxies will
decline by about a factor of 2.4. Assuming that the galaxies selected
for study here are a subset of this population, our numbers are in reasonable 
agreement with the predicted values. Unfortunately, the resolution of the 
simulations
by \citet{2001MNRAS.327L..10N} are too low to check whether the sizes of 
discs evolve self-consistently in this context.

We conclude that large disc galaxies have undergone substantial evolution
from $z=0.9$ to the present. The offset in the Tully-Fisher relation
is about $-1.1$ mag in the rest-frame B-band, and is related to an
overall increase in the central surface brightnesses of disc galaxies
and a decrease in rotation speed. We also find that the slope of the
Tully-Fisher relation, due to size constraints that inevitably result
from needing to obtain spatially-resolved rotation curves, cannot
meaningfully be estimated. In a subsequent paper (Barden et al. 2003),
we will discuss in detail the change in rotation speeds, disc masses,
disc scale lengths, and disc angular momenta of this sample of high
redshift galaxies relative to the local comparison sample, and make
quantitative comparisons with theoretical models of galaxy evolution.

\acknowledgments

We wish to thank the ESO OPC for the generous allocations of telescope
time and the Paranal staff for their expert support and assistance during
the observations.

\clearpage

\begin{deluxetable}{lcccccccrr}
\tabletypesize{\scriptsize}
\tablecaption{The high redshift sample \label{tbl-1}}
\tablewidth{0pt}
\tablehead{
\colhead{ID}&\colhead{z}&\colhead{r$_{25}$}&
\colhead{$\frac{a}{b}$}&\colhead{i}&\colhead{$A_{B}^{i}$}&
\colhead{$A_{B}^{g}$}&\colhead{m$_{B}$}&\colhead{M$_{B}^{rest}$}&
\colhead{v$_{rot}$}\\
\colhead{}&\colhead{}&\colhead{[kpc]}&
\colhead{}&\colhead{[deg]}&\colhead{[mag]}&
\colhead{[mag]}&\colhead{[mag]}&\colhead{[mag]}&
\colhead{[km s$^{-1}$]}\\
\colhead{(1)}&\colhead{(2)}&\colhead{(3)}&
\colhead{(4)}&\colhead{(5)}&\colhead{(6)}&
\colhead{(7)} &\colhead{(8)}&\colhead{(9)} &
\colhead{(10)}
}
\startdata
CFRS-00.0174&0.7838&12.9&3.58&79&0.11&0.14&23.22&-20.22$\pm 
$0.15&116$\pm $40\\
CFRS-00.0308&0.9704&17.8&1.33&43&0.06&0.13&23.72&-20.30$\pm $0.20&86$\pm 
$20\\
MS1054-1403&0.8133&40.7&3.21&76&0.54&0.15&20.74&-22.81$\pm $0.10&232$\pm 
$50\\
MS1054-1733&0.8347&14.9&1.88&60&0.22&0.15&22.97&-20.65$\pm $0.10&136$\pm 
$20\\
LDSS2-03.219&0.6024&13.9&1.73&57&0.18&0.34&23.25&-19.49$\pm 
$0.20&101$\pm $50\\
CFRS-03.0999&0.7049&30.3&3.53&78&0.62&0.42&21.45&-21.71$\pm 
$0.30&223$\pm $10\\
CFRS-03.1393&0.8554&22.6&4.23&83&0.69&0.42&22.26&-21.43$\pm 
$0.10&187$\pm $10\\
CFRS-03.1650&0.6341&19.6&2.78&72&0.43&0.42&22.33&-20.56$\pm 
$0.20&165$\pm $40\\
CFRS-22.0953&0.9787&23.9&2.41&68&0.34&0.28&22.87&-21.17$\pm 
$0.10&144$\pm $30\\
CFRS-22.1313&0.8173&23.9&2.92&74&0.47&0.28&22.29&-21.27$\pm 
$0.10&120$\pm $10\\
CN84-023&0.6389&25.4&1.47&49&0.12&0.17&20.78&-22.13$\pm $0.35&172$\pm $50\\
CN84-123&0.6776&20.3&1.98&62&0.24&0.17&21.48&-21.58$\pm $0.10&212$\pm $40\\
CSH96-68&1.5625&17.5&1.87&60&0.22&0.29&23.56&-21.73$\pm $0.35& 96$\pm $50\\
HDFS-0620\tablenotemark{a}&1.2850& 
3.6&1.49&49&0.12&0.12&28.00&-16.77$\pm $0.90&123$\pm $20\\
SA68-5155\tablenotemark{a}&1.0521&25.1&2.35&68&0.33&0.22&22.98&-21.26$\pm 
$0.10&160$\pm $50\\
CFRS-00.0137&0.9512&18.1&1.39&45&0.06&0.14&22.37&-21.60$\pm 
$0.10&237$\pm $40\\
CSH96-32&1.0215&13.1&1.49&49&0.12&0.30&22.85&-21.31$\pm $0.10&209$\pm $20\\
CSH96-74&1.3633&25.8&3.78&80&0.68&0.29&23.30&-21.63$\pm $0.10&112$\pm $30\\
CFRS-03.0776&0.8835& 7.6&1.22&36&0.06&0.42&23.69&-20.08$\pm 
$0.10&144$\pm $10\\
CFRS-03.1056&0.9392&13.8&1.50&50&0.13&0.42&22.16&-21.77$\pm 
$0.10&140$\pm $10\\
CFRS-03.1284&0.9393& 8.1&1.18&33&0.04&0.42&23.58&-20.36$\pm $0.10&167$\pm 
$20\\
CFRS-22.0599&0.8856&13.9&1.76&57&0.19&0.29&22.29&-21.48$\pm 
$0.10&159$\pm $30\\
\enddata

\tablenotetext{a}{galaxy observed by White et al. (ESO proposal ID:
63.O-0372(A))}

\tablecomments{(1) galaxy identification \citep{1995ApJ...455...50L,
1995MNRAS.273..157G, vandokkumthesis, 1998ApJ...497..188C,
1996AJ....112..839C} (2) redshift, (3) isophotal radius at
a surface brightness of 25 mag arcsec$^{-2}$, (4) ratio of
major over minor half axis, (5) inclination, (6) internal
extinction \citep{1985ApJS...58...67T}, (7) galactic extinction
\citep{1998ApJ...500..525S}, (8) total rest-frame B-band magnitude,
(9) total absolute rest-frame magnitude plus error (assuming $H_{0}=70.$, 
$\Omega _{M}=0.3$, $\Omega _{\Lambda }=0.7$), (10) rotation speed
corrected for inclination plus error}

\end{deluxetable}

\end{document}